# Modeling the Influence of Antifreeze Proteins on Three-Dimensional Ice Crystal Melt Shapes using a Geometric Approach

Jun Jie Liu[1,2], Yangzong Qin[1], Maya Bar Dolev[3], Yeliz Celik[1], J. S. Wettlaufer[4,5], and Ido Braslavsky[1,3]

[1] Department of Physics and Astronomy, Ohio University, Athens, Ohio
[2] School of Physical Science and Technology, Inner Mongolia University, Hohhot city, China
[3] The Robert H. Smith Faculty of Agriculture, Food & Environmental sciences, The Hebrew University of Jerusalem, Rehovot, Israel
[4] Department of Physics, Department of Geology and Geophysics, and Program in Applied Mathematics, Yale University, New Haven, Connecticut, USA
[5] NORDITA, Roslagstullsbacken 23, 10691 Stockholm, Sweden

**Abstract**

The melting of pure axisymmetric ice crystals has been described previously by us within the framework of so-called *geometric crystal growth*. Nonequilibrium ice crystal shapes evolving in the presence of hyperactive antifreeze proteins (hypAFPs) are experimentally observed to assume ellipsoidal geometries ("lemon" or "rice" shapes). To analyze such shapes we harness the underlying symmetry of hexagonal ice $I_h$ and extend two-dimensional geometric models to three-dimensions to reproduce the experimental dissolution process. The geometrical model developed will be useful as a quantitative test of the mechanisms of interaction between hypAFPs and ice.

**Introduction**

Understanding how the basic tenets of surface thermodynamics control crystal shapes is a major focus of condensed matter science. Using a ubiquitous material like ice as a test bed facilitates the advancement of these goals, while having immediate relevance to a wide range of applications. By independently controlling the temperature and pressure under which an ice crystal is grown, a range of equilibrium ice crystal shapes may be formed, from fully faceted surfaces to rough or rounded surfaces (*1, 2*).

It is rare to observe crystals in equilibrium; usually their shapes are probed in conditions of disequilibrium during which they are growing or melting, however slowly. Understanding the relationship between growth and equilibrium shapes (*3*) is basic to unraveling the kinetic processes on surfaces and their relationship to lower dimensional phase transitions, physisorption and chemisorption (2, 4). Equilibrium crystal shapes can be determined from the Wulff construction (5), which minimizes the surface energy for the fixed total volume of the sample. This geometrical construction requires as specified (or experimentally determined) the specific surface free energy $\gamma(\hat{N})$ as a function of crystallographic orientation relative to the surface normal $\hat{N}$. For clarity we note that $\gamma(\hat{N})$ for a liquid is the surface tension and hence is isotropic, independent of the surface normal $\hat{N}$, and single valued. One then performs the transform $\hat{N} \cdot \vec{r} = \lambda \gamma(\hat{N})$ where $\vec{r}$ is a radial vector from the origin to the crystal surface. The equilibrium crystal shape emerges by taking the interior envelope of the set of planes lying perpendicular to radial rays that intersect the polar plot of $\gamma(\hat{N})$. The overall size is characterized by $\lambda$, but the shape is independent of it (*3*). An equilibrium shape may be fully facetted (at low temperatures), it may be fully rough (at temperatures exceeding the roughening transitions for all of its facets) or both types of surface structure may coexist on a single crystal. Additionally, one might prepare an equilibrium crystal shape and use the Wulff construction to estimate the specific surface free energy $\gamma(\hat{N})$. However, due to the long relaxation times of typical materials, it is experimentally challenging to observe equilibrium crystal shapes. Notable successes include crystals of helium in contact with its superfluid parent phase, due to the negligible latent heat of fusion and large thermal conductivity (6), micron size crystals of metals (7), due to their small sizes and the ability to maintain ultra-high vacuum conditions over long periods of time, and ice crystals in contact with water in a well designed precision high pressure apparatus (1, 8).

Knowledge of the kinetic processes of molecular attachment, detachment, and surface diffusion and incorporation are fundamental to predicting the evolution of a crystal interface on length scales much larger than the lattice parameter. Of significant basic interest and broad applicability is to gain an understanding of the relationship between the macroscopic shape of a pure ice crystal and the microscopic behavior of water molecules. Moreover, the growth habits of ice can be influenced by impurities or additives such as antifreeze proteins (AFPs) that have specific affinity to certain crystallographic planes (9-13). Thus, quantitative studies of the non-equilibrium shapes of ice in the presence of AFPs may shed light on the kinetic processes underlying their action.

The driving force for both growth and dissolution is defined as the chemical potential difference between the parent phase and the solid, $\Delta\mu$. When the drive is sufficiently small that gradients in the rate limiting diffusion field possess a characteristic length that is large relative to the size of the crystal, then interfacial kinetic processes control the shape of the moving interface. Such a circumstance is common in a wide range of systems and settings and is modeled within the framework of *geometric crystal growth (1, 2, 14, 15)*. One can consider geometric crystal growth as a kinetically grounded kinematic approach. This is because the local interfacial velocity is specified based on the local crystallographic orientation $\hat{N}$, and on the kinetics understood to be associated with it, for a given $\Delta\mu$. Such approaches are successful in two-dimensional (2-D) growth having explained quantitatively the evolution of axisymmetric shapes (*1, 2*). Here, motivated by experiments involving AFPs, which substantially influence the interfacial kinetics and hence overall crystal shape, we extend geometric approaches to three dimensions and focus specifically on the shapes of ice crystals during dissolution. Note that while strictly speaking the shapes we model geometrically are dissolution shapes, here we use interchangeably dissolution and the more intuitive term melting. The distinction is irrelevant for either the actual shape or the method by which we model it.

The geometric approach explains the transient evolution of equilibrium forms containing both rough and faceted surfaces (14-16). Because the rough orientations lack an interfacial nucleation barrier whereas the faceted orientations grow by an activated process, the general growth process is one whereby the rough orientations grow out of existence and leave the shape dominated by slow growing facets. This is termed "global kinetic faceting". In treating the melting of an initially faceted shape, growth–melt asymmetries seen in ice crystals were captured (*1*). There, an apparent rotation of the principal facets by 30 degrees was explained by the underlying kinetic anisotropy of $V(\Delta\mu,\hat{N})$ due to step propagation. Pertaya *et al.* (12) observed similar experimental phenomena in AFP solutions, attributing it to the inhibition of crystal growth by AFPs bound to ice crystal surfaces. While the prism facets in pure ice in contact with water disappear above -16 °C and at pressures below 160 MPa (*1*), hypAFPs in solution created highly faceted hexagonal ice crystals viewed along the c-axis and ellipsoidal or "lemon-shaped" prismatic ice crystal faces at normal pressures and at temperatures near 0 °C (Figure 1). Modeling these complex three-dimensional shapes requires an extension of two-dimensional geometric models. Here, such an extension has been developed and allows us to simulate the observed hypAFP-modified shape evolution.

**Geometric Crystal Growth**

A small, spatially uniform and steady driving force characterizes near equilibrium growth or melting (*3*), both of which are described by geometric crystal growth (1-3, 14-16). By "small" it is meant that while the drive is large enough to change the shape of a crystal, if the drive is removed the shape relaxes to the equilibrium value on an experimentally observable time scale. As in the Wulff construction, where the equilibrium shape is determined by the orientation dependence of the specific surface free energy $\gamma(\hat{N})$, in geometric growth or melting the shape is determined by the orientation dependence of the local interfacial velocity $V(\Delta\mu,\hat{N})$, itself dependent upon the growth drive. For 2-D growth one describes the surface itself by a time-dependent vector $\vec{C}[x(u,t),y(u,t)]$, in Cartesian coordinates $x(u,t), y(u,t)$ depending on an arc-length parameter, u, related to the arc-length $s$ by the metric $\sqrt{(\partial x/\partial u)^2 + (\partial y/\partial u)^2} \equiv |\partial\vec{C}/\partial u|$ in the usual manner; $ds = |\partial\vec{C}/\partial u|du$. Whence, for an inward pointing unit normal $\hat{N}$ and an anisotropic velocity function $V(\Delta\mu,\hat{N})$, the surface evolves according to

$$\left(\frac{\partial \vec{C}(u,t)}{\partial t}\right)_u = -V\hat{N}, \qquad [1]$$

where we suppress the explicit representation of the Cartesian two-vector because the arc-length convention implies reference to a point on an evolving two-dimensional surface. Note that the time derivative is taken at fixed arc length. An equivalent means for dealing with interfacial kinetics as a function of $\hat{N}$ is to characterize that direction in terms of the angle $\theta$ that the surface tangent makes with the positive $x$–axis, leading to $V(\Delta\mu,\theta)$. For a given velocity function, the evolution of an initial shape is given by the solution to equation [1] using the method of characteristics *(15)*.

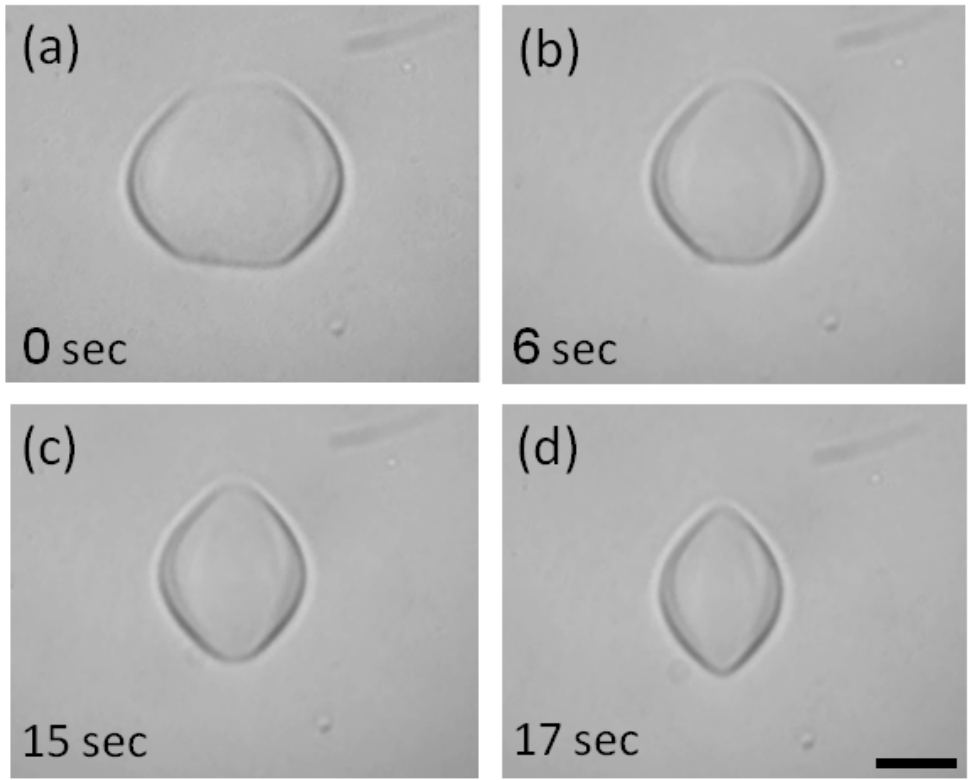

**Figure 1.** Melting of an ice crystal in the presence of *Tenebrio molitor* AFP. (a)–(d) melting process observed in nanoliter osmometer experiments (17). The formation of the ellipsoidal or "lemon" shape is evident viewed perpendicular to the c-axis. The scale bar is 10 μm and the times are shown in the figures.

Independent of whether one is describing the evolution of a two-dimensional or three-dimensional, as long as the conditions of geometric growth are met, the global evolution of an initial crystal is governed by the physics embodied in a local velocity function $V$. However, a two-dimensional surface bounding a three-dimensional crystal must be parameterized by two arc-lengths modifying equation [1] as follows,

$$\left(\frac{\partial \vec{C}(u,v,t)}{\partial t}\right)_{u,v} = -V\hat{N}, \qquad [2]$$

where the arc-length parameters $u$ and $v$ correspond to the longitudinal and latitudinal positions on the evolving crystal surface. Thus, by parity of reasoning with the two-dimensional case, the arc-lengths $u$ and $v$ carry the same information as do the angles $\theta$ and $\varphi$ that define the tangent plane perpendicular to the surface normal $\hat{N}$. The tangent plane is defined by the unit surface tangent $\hat{T}$, whose angle $\theta$ is measured from the positive $x-$axis, and the unit binormal $\hat{B}$, perpendicular to $\hat{T}$, which has a projection onto the equatorial plane taking an angle $\varphi$ relative to the $x-$axis (figure 2).

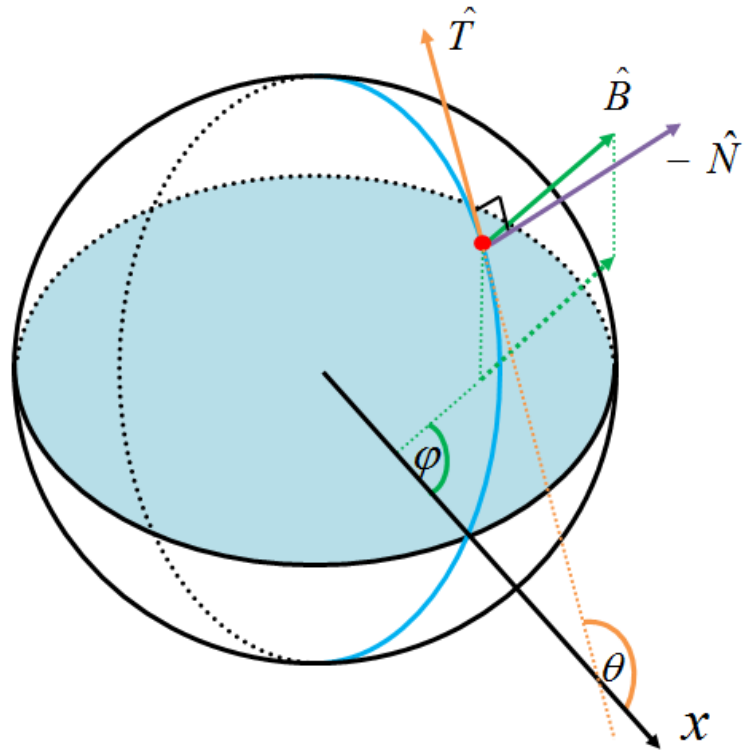

**Figure 2.** Evolution of a single point on a crystal surface with a normal $\hat{N}$, tangent $\hat{T}$, and binormal $\hat{B}$.

Whence, the local normal velocity has the following dependences; $V(\Delta\mu,\theta,\varphi)$, it is intuitive that a unique correspondence between $u$, $v$ and $\theta$, $\varphi$ exists globally for continuous convex shapes. When a discontinuity forms at a corner, the transformation is piecewise valid on the continuous regions separating corners. A surface that becomes faceted, and hence has zero curvature, must be treated separately such as described in (14), the results of which we extend presently, by rewriting equation [2] as

$$\left(\frac{\partial \vec{C}(\theta,\varphi,t)}{\partial t}\right)_{\theta,\varphi} = -V(\Delta\mu,\theta,\varphi)\hat{N} + \left(\frac{\partial V}{\partial \theta}\right)_{\varphi,t}\hat{T} + \left(\frac{\partial V}{\partial \varphi}\right)_{\theta,t}\hat{B}, \qquad [3]$$

described in more detail in the Appendix. Due to the fact that, for a given driving force, $V(\Delta\mu,\theta,\varphi)$ depends solely on the orientation of the normal, the solution $\vec{C}(\theta,\varphi,t)$ of equation 3 is given along straight characteristic lines along which the normal is preserved;

$$\vec{C}(\theta,\varphi,t) = \vec{C}_o(\theta,\varphi) + \left[-V(\Delta\mu,\theta,\varphi)\hat{N} + \left(\frac{\partial V}{\partial \theta}\right)_{\varphi,t}\hat{T} + \left(\frac{\partial V}{\partial \varphi}\right)_{\theta,t}\hat{B}\right]t. \qquad [4]$$

All orientations on the initial surface $\vec{C}_0(\theta,\varphi)$ move along characteristic lines and at each time $t$ the locus of all points $\vec{C}(\theta,\varphi,t)$ associated with them define the crystal shape. When two characteristics meet, a discontinuity in the surface slope can occur which can be described as a shock (14, 16). The trajectory of these shocks has been analyzed in the two-dimensional case (14). Simply reversing time allows for the treatment of melting.

Now we construct a simple velocity function $V(\Delta\mu,\theta,\varphi)$ geared towards our interests in the anisotropic melting shape of an ice crystal viz.,

$$V(\Delta\mu,\theta,\varphi) = V_f(\Delta\mu)\xi(\theta) + V_r(\Delta\mu,\theta,\varphi)[1-\xi(\theta)], \qquad [5]$$

where $V_f$ is the velocity normal to faceted surfaces, such as the basal plane, and $V_r$ is that along rough surfaces, such as all prismatic planes which here become rough over time. The function $\xi(\theta)$ captures the transition from faceted to rough growth along the surface and is periodic with a period of $n$ and thus continuously varies between zero and one; $0 \le \xi(\theta) \le 1$. For simplicity, we chose $\xi(\theta) = \cos^m(n\theta/2)$, where $m$ is an even integer. Motion of a molecularly rough interface is not hindered by activated processes, such as is the case at a facet, because for growth or melting there is no nucleation barrier; the surface can be thought of heuristically as a liquid-vapor interface. Whence, $V_r$ is a relatively simple function of the growth drive $\Delta\mu$, which under experimental conditions is either proportional to the undercooling at fixed pressure or the overpressure at fixed temperature (1, 8, 16). Using the intuition gleaned from both growth and melting studies of pure materials (1, 15, 16), we extend the linear response regime to the two-dimensional surface as follows;

$$V_r(\Delta\mu,\theta,\varphi,) = c_r \Delta\mu \left[1 + \cos^p\left(\frac{n_\theta \theta}{2}\right)\right]\left[1 + \cos^q\left(\frac{n_\varphi \varphi}{2}\right)\right], \quad [6]$$

in which the linear response is shown in the first term with the growth drive scaled by a constant $c_r$. Moving along a surface from rough orientations towards vicinal orientations, the normal motion of an interface is influenced both by detachment of molecules along with the migration of surface molecules away from the facets (1, 15). This behavior is captured by the two angle dependent terms, in which $p$ and $q$ are even integers, and $m \geq p,q$.

In the experiments reported here and elsewhere (1, 8, 16) it is seen that ice crystal shapes are two-fold symmetric parallel to the basal plane and six fold symmetric about the c-axis, and hence we take the simplest form of equation 6 that describes the symmetry of the crystal; $p = q = 2$, $n_\theta = 2$ and $n_\varphi = 6$ (Figure 3). Under the conditions of geometric growth or melting, wherein the driving force is weak, $V_f \ll V_r$, and hence for example the advance of the basal plane is much slower than that of the prismatic planes in experiments.

Given an initial shape, each point on the crystal advances according to the orientation of the surface. Depending on the initial shape, trajectories of two points may collide in a shock event (14, 16), according to the rubric of the method of characteristics. When a shock occurs, the participating characteristic orientations are eliminated from the crystal surface (14). The right hand side of equation 4 displays the wide range of possibilities, depending on the initial shape and the detailed form of $V(\Delta\mu,\theta,\varphi)$, for shock formation.

**Results**

Experiments on ice crystal growth in the presence of hypAFPs revealed highly faceted crystals in the basal plane (17). AFPs inhibit ice crystals from growing within a certain temperature range, thereby decreasing the freezing point below the equilibrium melting temperature (9). When crystals grown in solutions containing AFPs are continuously cooled and the magnitude of the freezing point is exceeded, sudden and rapid growth occurs due to the magnitude of supercooling imposed. In contrast, melting can be controlled using relatively small temperature gradients, and shape evolution is readily observed under a microscope (18). We used the three-dimensional model described above to simulate the melting process and to demonstrate how ice crystals melt in hypAFP solutions.

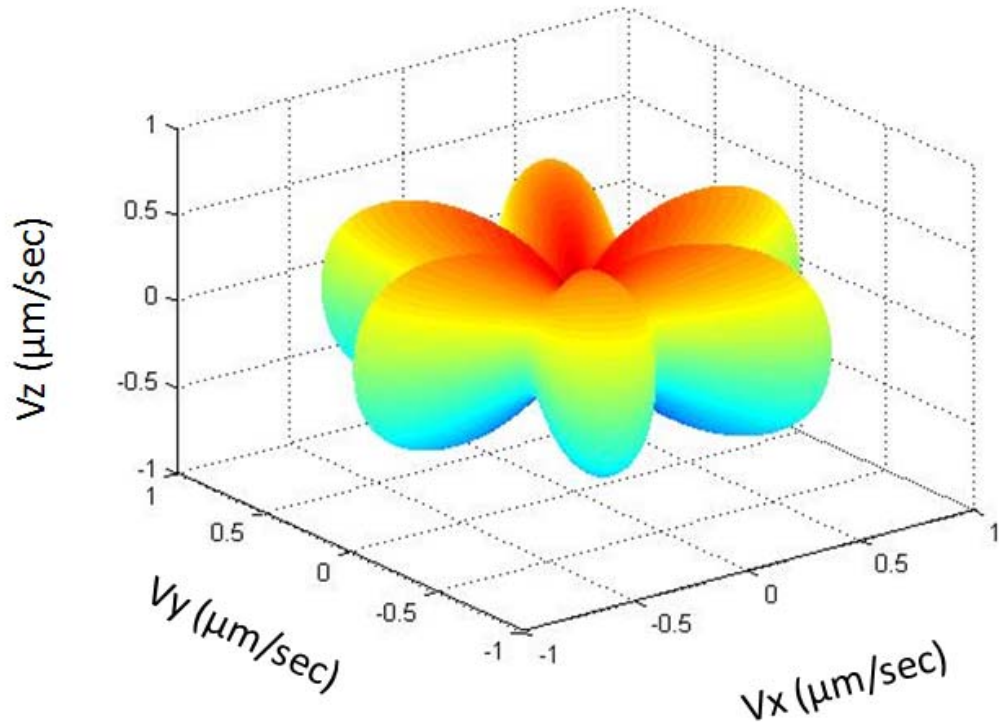

**Figure 3.** The three-dimensional velocity structure shown in terms of the cartesian velocity components (Vx, Vy, Vz). The velocity parameters are $c_r\, \Delta\mu = 0.48$ μm/sec and $V_f = 0.08$ μm/sec. The reversal of time to treat dissolution is equivalent to reversing the sign of this velocity as was done for the simulations shown in figures 4 and 5. The color from dark blue to red indicates the variation of the angle θ from 0 to 180 degrees.

The parameters were chosen to fit the simulation results to the experimental results of melting of crystals in *Tenebrio molitor* AFP (*Tm*AFP) solutions. We note that in such a manner comparison with other AFP solutions will offer future insight into system specificity versus generality of this approach.

Figure 4 shows model results of a dissolving ice crystal viewed parallel to the c-axis. As time progresses, the initially rounded shape evolves into a hexagonal structure and the faceted basal plane diminishes in overall area. This is a transformation commonly observed in nanoliter osmometer experiments with *Tm*AFP solutions (17), distinct from the pure case in which the basal facet is maintained (8).

Figure 5 displays the evolution perpendicular to the c-axis. The oblate spheroidal, or "drum-like", initial shape forms ridges during melting. The relatively slow ablation normal to the basal plane creates a dynamically faceted oblate spheroid of the opposite aspect ratio that we refer to as a truncated lemon. Such an evolution is seen in the experimental observations shown in Figure 1. Clearly, the morphology of these crystal surfaces is not arbitrary, but arises from a general set of underlying rules. Finally, as is seen particularly well in Figures 5 (c) and (d), regions of the crystal become faceted during ablation and hence their local curvature must decrease.

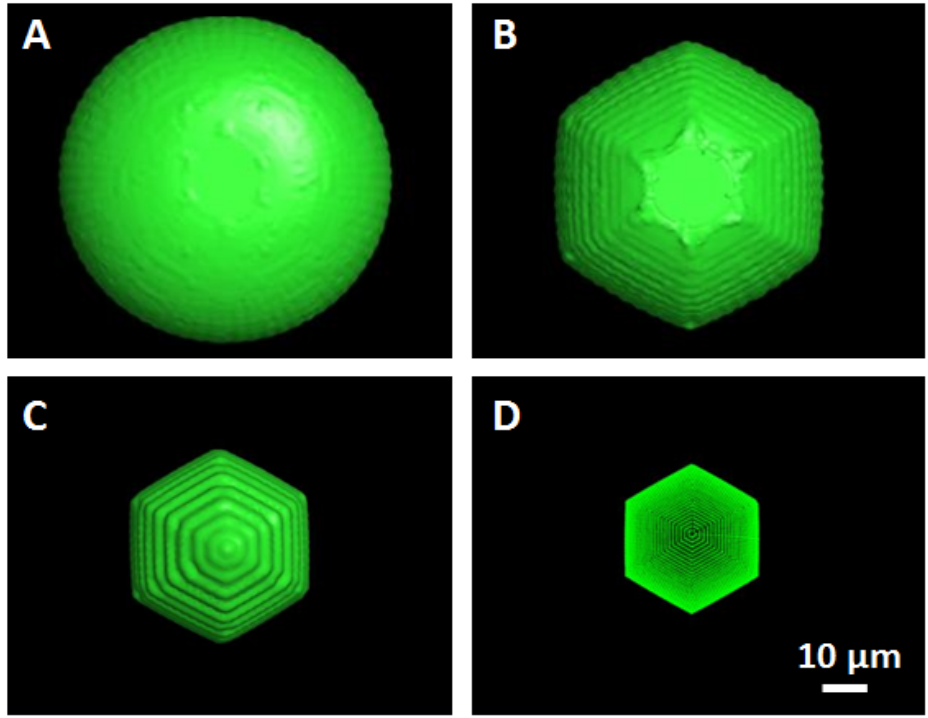

**Figure 4.** Computer generated model of a melting shape that results from the velocity profile that is described in Figure 3, viewed along the c-axis starting from an oblate spheroidal shape. A view along the c-axis reveals the shapes at four points in time (a) 0 sec (b) 8 sec (c) 18 sec (d) 22 sec. The clear hexagonal shape of the 3-D model is in agreement with experiments (17).

The evolution of the local curvature on certain regions of a crystal surface in which only one of the principal orientations is important can be analyzed in ostensibly the same manner as in 2-D (1, 14-16). For example, this might be treated by fixing the angle $\varphi$ associated with the binormal $\hat{B}$ and solving a local equation for the curvature $\kappa_\theta$ as a function of the angle $\theta$ associated with the tangent $\hat{T}$ for a one-dimensional boundary viz.,

$$\frac{\partial \kappa_\theta}{\partial \tau} = -\kappa_\theta^2 \tilde{V}_\theta .  \qquad [7]$$

In this equation $\partial/\partial\tau$ is taken at constant $\theta$, the so-called "velocity stiffness" $\tilde{V}_\theta \equiv V_\theta + \left(\frac{\partial^2 V_\theta}{\partial \theta^2}\right)$ is determined at fixed $\varphi$, which is, *mutatis mutandis,* the same as in (15). Whence, for fixed $\varphi$, the evolution of the curvature as a function of $\tau$ from the initial value $\kappa_{\theta I}$ is

$$\kappa_\theta = \frac{\kappa_{\theta I}}{1 + \kappa_{\theta I} \tilde{V}_\theta \tau} . \qquad [8]$$

We note however that any local axisymmetry of the two-dimensional surface that allows this heuristic interpretation of the observed curvature decrease will in general be limited to a finite time interval. After such time it is possible that the curvature evolution in the other direction will intervene, as well as issues associated with the formation of shocks associated with regions on the surface with negative surface stiffness as discussed above and in (1, 14-16).

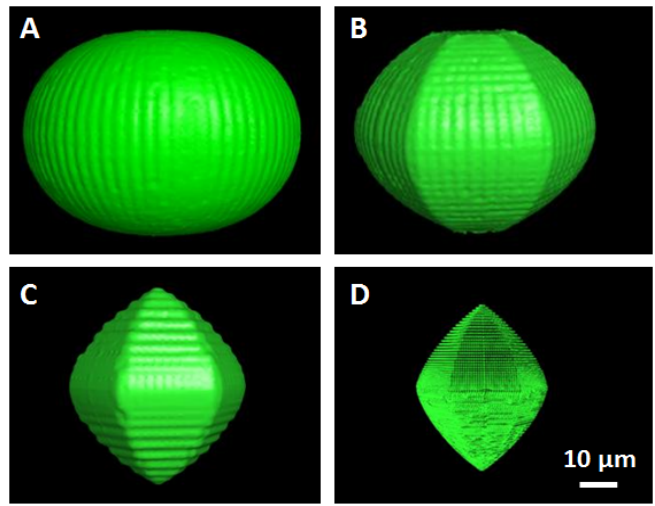

**Figure 5.** Computer generated model of the melting shape in figure 4 viewed along the a-axis at (a) 0 sec (b) 8 sec (c) 18 sec (d) 22 sec. The initial "drum" shape evolves into a "faceted lemon" which maintains the same form as it disappears, as is seen in the experimental observations shown in Figure 1.

**Conclusions**

We have extended two-dimensional *geometric models* for crystal growth and melting to three-dimensions and have examined the implications for the problem of ice crystals melting (dissolving) in solutions of hyperactive antifreeze proteins (hypAFPs). It is found that, consistent with observations of ice crystals in solutions of *Tm*AFP, initially rounded crystals melted to form stable faceted "lemon shapes". The detailed morphology of such shapes is expected to depend on environmental conditions, such as the thermal hysteresis promoted by the AFP, the concentration of AFP, and the driving force $\Delta\mu$, all of which affect the underlying velocity function $V(\Delta\mu,\theta,\varphi)$. Indeed, having demonstrated the ability of geometric theories to capture detailed morphological changes in this complex system, the next stage of research aims at systematically testing the environmental influences on $V(\Delta\mu,\theta,\varphi)$. In this manner, refinement of both model structure and protein-ice interactions is possible. Importantly, such an approach provides quantitative insight concerning how macroscopic shapes are controlled by the mechanisms of dissolution and melting as they are modified by the presence of AFPs.

**Acknowledgements**

This work has been supported by the National Science Foundation (NSF) under Grant No. CHE-0848081, the Condensed Matter and Surface Science program at Ohio University (CMSS), the Israel Science Foundation (1279/10 and 1281/10), the Marie Curie International Reintegration Grant (256364) and by the European Research Council (281595). JJL thanks the National Natural Science Foundation of China and the Science (31160188) and Technology Foundation of Ministry of Education of China (211030) for their research support. JSW thanks the Helmholtz Gemeinschaft Alliance "Planetary Evolution and Life," the Wenner-Gren Foundation, the John Simon Guggenheim Foundation, and the Swedish Research Council for generous support of this research.

**Appendix A: Derivation of the three-dimensional model**

The approach here is a straightforward extension of Tsemekhman and Wettlaufer (14) and thus we provide here only the major waypoints sufficient to allow the reader to follow the logic. The evolution equation [2] is

$$\left(\frac{\partial \vec{C}(u,v,t)}{\partial t}\right)_{u,v} = -V\hat{N}$$

[A1]

where the arc-length parameters $u$ and $v$ correspond to the longitudinal and latitudinal positions corresponding to angles $\theta$ and $\varphi$ that define the tangent plane perpendicular to the surface normal $\hat{N}$. The plane is collinear with the unit surface tangent $\hat{T}$, whose angle $\theta$ is measured from the positive $x-$ axis, and the unit binormal $\hat{B}$, perpendicular to $\hat{T}$, which has a projection onto the equatorial plane taking an angle $\varphi$ relative to the $x-$ axis (figure 2). In terms of $u$ and $v$ we write the usual arc lengths as $ds_u = \left|\partial \vec{C}/\partial u\right|_v du$ and $ds_v = \left|\partial \vec{C}/\partial v\right|_u dv$, and with the use of

$$\left(\frac{\partial}{\partial t}\right)_{\theta,\varphi} = \left(\frac{\partial}{\partial t}\right)_{u,v} - \left(\frac{\partial \theta}{\partial t}\right)_{u,v}\frac{\partial}{\partial \theta} - \left(\frac{\partial \varphi}{\partial t}\right)_{u,v}\frac{\partial}{\partial \varphi},$$

[A2]

a reparameterization of the surface is generalized from (14) as

$$\left(\frac{\partial \vec{C}(\theta,\varphi,t)}{\partial \theta}\right)_\varphi = \frac{\partial \vec{C}(\theta,\varphi,t)}{\partial s_u}\frac{\partial s_u}{\partial \theta} = \frac{1}{\kappa_\theta(\theta,t)}\hat{T}.$$

[A3]

$$\left(\frac{\partial \vec{C}(\theta,\varphi,t)}{\partial \varphi}\right)_\theta = \frac{\partial \vec{C}(\theta,\varphi,t)}{\partial s_v}\frac{\partial s_v}{\partial \varphi} \equiv \frac{1}{\kappa_\varphi(\varphi,t)}\hat{B},$$

[A4]

Here $\kappa_\theta(\theta, t)$ is a compact form of the curvature as a function of $\theta$ and t for fixed $\varphi$ and the same nomenclature follows, the obvious changes being made, for $\kappa_\varphi(\varphi, t)$. Combining equations [A2], [A3], and [A4] yields

$$\left(\frac{\partial \vec{C}(\theta,\varphi,t)}{\partial t}\right)_{\theta,\varphi} = -V(\theta,\varphi)\hat{N} - \frac{1}{\kappa_\theta(\theta,t)}\left(\frac{\partial \theta}{\partial t}\right)_{u,v}\hat{T} - \frac{1}{\kappa_\varphi(\varphi,t)}\left(\frac{\partial \varphi}{\partial t}\right)_{u,v}\hat{B} \quad . \qquad [A5]$$

The appropriate generalization of equation (4) from (14) gives the pair

$$\left(\frac{\partial \theta}{\partial t}\right)_{u,v} = -\frac{\partial V}{\partial s_u} \equiv -\kappa_\theta \frac{\partial V}{\partial \theta} \quad \text{and} \quad \left(\frac{\partial \varphi}{\partial t}\right)_{u,v} = -\frac{\partial V}{\partial s_v} \equiv -\kappa_\varphi \frac{\partial V}{\partial \varphi} \quad . \qquad [A6]$$

Combining equations [A5] and [A6] yields

$$\left(\frac{\partial \vec{C}(\theta,\varphi,t)}{\partial t}\right)_{\theta,\varphi} = -V(\Delta\mu,\theta,\varphi)\hat{N} + \left(\frac{\partial V}{\partial \theta}\right)_{\varphi,t}\hat{T} + \left(\frac{\partial V}{\partial \varphi}\right)_{\theta,t}\hat{B} \quad , \qquad [A7]$$

which is equation 3, the solution to which is equation [4].

**References**

1. Cahoon A, Maruyama M, & Wettlaufer JS (2006) Growth-melt asymmetry in crystals and twelve-sided snowflakes. *Phys. Rev. Lett.* **96**(25). DOI: 10.1103/PhysRevLett.96.255502.
2. Wettlaufer JS (2001) Dynamics of ice surfaces. *Interface Sci.* **9**:115-127. DOI: 10.1023/A: 1011287217765.
3. Elbaum M & Wettlaufer JS (1993) Relation of growth and equilibrium crystal shapes. *Phys. Rev. E* **48**:3180. DOI: 10.1103/PhysRevE.48.3180.
4. Gilmer GH & Bennema P (1972) Simulation of crystal-growth with surface diffusion. *J. Appl. Phys.* **43**(4):1347-&. 10.1063/1.1661325.
5. Wulff G (1901) Zur Frage der Geschwindigkeit des Wachstums und der Auflösung der Kristallflächen. *Z. Krist* **34**:449-530.
6. Lipson SG (1987) Helium Crystals. *Contemp. Phys.* **28**(2):117 - 142. DOI: 10.1080/00107518708223690.

7. Heyraud JC & Metois JJ (1980) Establishment of the Equilibrium shape of metal crystallites in a foreign substrate: Gold on Graphite. *J. Cryst. Growth* **50**:571 - 574. DOI:10.1016/0022-0248(80)90112-8.
8. Maruyama M (2011) Relation between growth and melt shapes of ice crystals. *J. Cryst. Growth* **318**: 36-39 DOI: 10.1016/j.jcrysgro.2010.10.076.
9. Yeh Y & Feeney R (1996) Antifreeze proteins: Structures and mechanisms of function. *Chem. Rev.* **96**(2):601-617. DOI: 10.1021/cr950260c.
10. Strom CS, Liu XY, & Jia Z (2004) Antifreeze protein-induced morphological modification mechanisms linked to ice binding surface. *J Biol Chem* **279**(31):DOI:32407-32417. 10.1074/jbc.M401712200.
11. Strom CS, Liu XY, & Jia Z (2005) Why does insect antifreeze protein from Tenebrio molitor produce pyramidal ice crystallites? *Biophys J* **89**(4):2618-2627. DOI:10.1529/biophysj.104.056770.
12. Pertaya N, Celik Y, DiPrinzio CL, Wettlaufer JS, Davies PL, & Braslavsky I (2007) Growth-melt asymmetry in ice crystals under the influence of spruce budworm antifreeze protein. *J. Phys-Cond. Matt.* **19**:412101. DOI: 10.1088/0953-8984/19/41/412101.
13. Wathen B, Kuiper M, Walker V, & Jia ZC (2003) A new model for simulating 3-D crystal growth and its application to the study of antifreeze proteins. *Journal of the American Chemical Society* **125**(3):729-737. 10.1021/ja0267932.
14. Tsemekhman V & Wettlaufer JS (2003) Singularities, Shocks, and Instabilities in Interface Growth. *Stud. Appl. Math.* **110**:221-256. DOI: 10.1111/1467-9590.00238.
15. Wettlaufer JS, Jackson M, & Elbaum M (1994) A geometric model for anisotropic crystal growth. *J. Phys A: Math. Gen* **27**:5957-5967. DOI: 10.1088/0305-4470/27/17/027.
16. Maruyama M, Kuribayashi N, Kawabata K, & Wettlaufer JS (2000) Shocks and curvature dynamics: A test of global kinetic faceting in crystals. *Phys. Rev. Lett.* **85**(12):2545-2548. DOI: 10.1103/PhysRevLett.85.2545.
17. Bar M, Celik Y, Wettlaufer JS, Davies PL, & Braslavsky I (2012) New insights on ice growth and melting modifications by antifreeze proteins. *Journal of the Royal Society Interface, in press*.
18. Bar M, Celik Y, Fass D, & Braslavsky I (2008) Interactions of beta-helical antifreeze protein mutants with ice. *Crystal Growth & Design* **8**(8):2954-2963. DOI: 10.1021/cg800066g.